**Critical Need for a National Initiative in Low Temperature Plasma Research**

Philip Efthimion[1], Igor Kaganovich[1], Yevgeny Raitses[1], M. Keidar[2], Hyo-Chang Lee[3], Mikhail Shneider[4], R. Car[4]

[1]*Princeton Plasma Physics Laboratory, Princeton University, Princeton NJ 08543, USA*
[2]*George Washington University, Washington DC 20052, USA*
[3]*Korea Research Institute of Standards and Science, Daejeon 34113, Republic of Korea*
[4]*Princeton University, Princeton NJ 08540, USA*

*Email: pefthimi@pppl.gov*
*Area: Discovery Plasma Science (DPS)*

**Goals of Initiative**
To initiate a national program in Low Temperature Plasma (LTP) to take advantage of the research opportunities of 3 rapidly growing areas (nanomaterial plasma synthesis, plasma medicine, microelectronics). The main theme is to achieve a fundamental understanding of Low Temperature Plasmas as they are applied to these different applications. This understanding will allow U.S. industry to meet the challenges of international competition.

**Programmatic Benefit**
The programmatic benefit of this initiative is that a fundamental understanding of the plasmas as they are applied to the areas of nanomaterial plasma synthesis, plasma medicine, and microelectronics will be developed. This will eventually replace the current Edisonian approach to applications research that is costly and inefficient. This is a major challenge because Low Temperature Plasmas are complex. They have low ionization and therefore neutrals, electrons, and ions interactions need to be understood. In all three of these application areas plasma chemistry plays a major role. In many cases the plasmas are not in equilibrium, not Maxwellian, and radiation effects are important. Furthermore, at low pressures the plasma mean free path or energy relaxation length is longer than the system size. Therefore, the boundary conditions impact the plasma volume of the system.

More importantly the initiative addresses the critical needs of the U.S. microelectronics industry, see white paper by microelectronics industry leaders [1]. The research needs for microelectronics have rapidly changed over the last 5 – 10 years. Where the US once dominated in the production of semi-conductor devices, now South Korea leads the world. Specifically, Samsung leads Intel in sale of such devices. Furthermore, there are concerns of state sponsored initiatives by China to modernize its economy, increase productivity, and use innovation to create economic growth. The initiative Made-In-China (MIC) 2025, looks to use innovation for future growth. In particular, China is targeting key industries to provide innovation including the microelectronics industry. An example of China becoming the leader of high-value manufacturing is their semiconductor purchases. In 2003 China was similar to the US (19.4% US, 18.5% China) in the purchase of semiconductors for the products it sold, but in 2019 the US has fallen to 11.9% and China dominates by consuming 60.5% of the world market [2]. These semiconductors are in the products China manufactures and sells to the world. *The 2025 initiative has its goal to move China from being the largest consumer of* *semiconductors to becoming the largest manufacturer of semiconductors.* Microelectronics is nearly a $2 T/year industry.

The US program should invigorate LTP research at the universities and national labs into a sustained program to develop plasma manufacturing techniques and understanding of the plasma conditions that are the basis of these manufacturing techniques. This joint partnership should not exist just for the short period of time needed to meet the immediate international challenge but should be sustained to continue to promote innovations and maintain our leadership of the field. At the moment, DOE's budget in LTP is on the order of $4M/year. NSF does not support LTP with any focused programs. This is in contrast to the $Bs invested by industry for their internal research and the gross sales of the microelectronics sector of $2 T/year. The government is overlooking this threat to the US economy, military superiority, and US jobs. The 2010 Decadal Report on Plasma Science recognized the lack of sustained support and lack of stewardship for Low Temperature Plasma research. The modest response of the government has been is insufficient to meet the international challenge. The size of the current market warrants a government program ($100 M/year) in LTP to promote and protect US interests similar to the programs set up to protect medical science and pharmaceuticals industries by the NIH (>$20 B/year); exascale computing (>$500 M) for an industry that is expected to emerge shortly, but currently does not generate substantial sales; and material science by BES of OS ($1.1 B/year). This initiative should not be funded at the expense of other important plasma application (e.g. domestic fusion energy research, ITER).

While plasma medicine is a new scientific field there are several established international research groups working on the frontline. For instance, in Germany INP Greifswald University is supported at multimillion level by German Government leading to INP leadership in cold plasma medical devices development, and standardization acceptance by medical community. In South Korea, Kwanwoon University is developing a cancer therapy based on cold plasma. They focus on biological effects as well as on engineering aspects of cold plasma devices. South Korea government support is at the level of $25m over 5 years. Similarly in Japan, Nagoya University group of Prof. M. Hori is developing plasma sources and performing in vivo studies and is well supported by Japanese government. *In summary, there is coordinated research effort supported by governments in Germany, Korea and Japan in plasma medicine area, while there is no such program in US.*

**Description of the Initiative**
Long-term investments on all three areas of Low Temperature Plasma: plasma synthesis of nanomaterials, plasma medicine, and microelectronics.

Specifically:

**1. Nanomaterial plasma synthesis**
LTP can be used to produce metal nanoparticles, for example, by decomposing a metal-containing gas, or to produce nanoparticles of the IV,V-group of nonmetallic elements of the periodic table (C, B, Si, N etc.). Nanoparticles are synthesized by either decomposing background gases or ablating electrodes containing these elements. The plasma environment offers a number of attractive properties that allow for the generation of nanoparticle materials that are hard to produce by other means. The negative nanoparticle charge in plasmas suppresses particle coagulation and enables the synthesis of highly monodisperse nanoparticles. The negative charge also confines nanoparticles in the reactor, very different from neutral gas synthesis, in which particles are often quickly lost to the reactor walls due to diffusion. These unique capabilities of LTP have already enabled technological advances such as the generation of nanometer-sized metal particles, diamonds, stable silicon-nanocrystal anodes for lithium ion batteries, high-mobility transparent conductive oxide films based on zinc oxide nanocrystals etc. [3].

Low Temperature Plasma that create nanomaterials can operate at temperatures near 10,000 degrees K and provide high fluxes of chemically active species that can trigger new chemical pathways for efficient nucleation and growth of nanomaterials with specific compositions. Even though plasma synthesis of nanomaterials has significantly higher growth rates compared to Chemical Vapor Deposition, its lack of selectivity limits its adaptation [4].   High selectivity requires developing new more sophisticated fabrication conditions beyond the conventional plasma approaches. For example, Kim et al. [5] manufactured BN nanotubes (NT) using a RF plasma with addition of hydrogen.   The presence of hydrogen is believed to promote the formation of BN precursors such as NH and BH radicals and greatly enhance production of BN nanotubes [5]. This is just one of examples of importance of plasma chemistry in formation of nanomaterials. These results predict possible new mechanisms for high temperature synthesis methods such as plasma and lasers, and suggest the possibility of controlling the synthesis process and the types of synthesized nanostructures through the addition of hydrogen.

There are a number of different proposed growth mechanisms for the synthesis of Boron Nitride (BN) nanostructures. Using quantum-classical Molecular Dynamics (MD), the synthesis of BNNTs and other BN nanostructures was simulated [6,7,8]. BN molecules can form BN cages without the presence of a boron NP to act as a catalyst at a temperature of 2000K, which is within the temperature range of the proposed synthesis region in the plasma torch experiments by direct self-organization of BN diatomic molecules. It was also found that BN nanostructures can be synthesized by bombarding a boron nanoparticle with BN diatomic molecules or nitrogen atoms [8]. However, if hydrogen-rich molecules such as $NH_3$ or HBNH are used as a feedstock, stable two-dimensional nanoflake structures are only formed [8]. In addition, BN nanotubes growth was demonstrated from a short BN nanotube template at 2500 K [8]. However, how this template can be created in practice is not clear.

Additional MD simulations [7] seem to suggest that for formation of BNNT a small droplet of boron in the temperature range ~ 2000-2500 K has to absorb enough nitrogen to form a boron nitride cap. Further growth of both boron droplet (due to boron flux onto the droplet) and BNNT (due to nitrogen flux onto the boron droplet) can occur as soon as BN cap is formed. The length of BNNT will be determined by how much nitrogen flux can incorporate into BNNT root, similar to growth of carbon nanotubes (CNT).

However, these MD simulations have been performed with pre-defined gas composition; in particular, pure BN molecules gas and pure borazine gas were separately considered. In practice, the feedstock material is introduced into plasma, evaporates, and is subsequently transported to the reaction zone. In this zone, under definite temperature conditions, boron starts to condense and nanotubes form. In this process, composition of the gas mixture cannot be arbitrary prepared; it is determined by chemical reactions between the gas species. What can be controlled is feedstock material, pressure and temperature within the system and buffer gas composition. Molecular dynamics simulations are not capable of modeling long timescale processes of material evaporation, cooling and condensation; different method should be applied to determine actual gas mixture composition at the nanoparticles growth [9,10].

Other important synthesis applications include synthesis of 2D materials and thin films, diamond-like materials, including diamond defects for quantum computing, and porous materials to name a few [11]. In all these applications plasma is often used for synthesis, and plasma properties are poorly understood.

Additional funding is needed to understand and develop reliable plasma nano-material synthesis techniques with high selectivity for all materials. To achieve this it is important to have a large array of advanced plasma diagnostics. The first steps towards in situ diagnosis of the nucleation and growth of nanoparticles in plasma have already been completed (see, for example in Ref. [12]). The next steps should include diagnosing the particle charging and structure of nanoparticles in situ. *Integrated modeling including molecular dynamics, thermodynamics, fluid modeling of plasma composition, should be performed in a self-consistent manner and validated by extensive experimental data obtained with in situ diagnostics to achieve understanding of synthesis and selectivity.*

## 2. Plasma medicine

Plasma medicine research is just in its infancy but is a rapidly growing field of medical treatment, with the number and type of medical applications growing annually, such as dentistry, cancer treatment, wound healing, antimicrobial and surface sterilization. Plasma medicine is a new, largely unexplored multi-disciplinary scientific field that integrates physics, chemistry, engineering and biology [13,14,15,16,17,18,19,20,21,22]. It focuses on the biomedical technology based on interaction of cold atmospheric plasma with cells alone and within the context of living tissue. Controlling non-thermal plasma interaction with biological objects (biological liquids, cells, tissues, food, plants, agricultural products) is one of the most important aspects of the plasma application in biomedical field. Furthermore, LTP promotes muscle and blood vessel regeneration, and osteointegration is also being investigated. With the rapid increase in new medical applications, there is development of new devices and systems for treatment, and wider clinical applications. There is a wide range of plasma sources that allow customization of specific applications. Plasma medicine is a powerful tool in medical research. Besides developing plasma medical applications, additional funding is required to understand the mechanisms for how plasma triggers biological processes to heal wounds, kill cancer cells preferentially, and kill bacteria [20].

## 3. Microelectronics

What is needed is a joint government-private sector research initiative to develop the techniques required to make micro-chips utilizing 10 nm lithography and meet the challenge of international mobilization to also develop alternative techniques, see Appendix for more information. In semiconductor processing and equipment, there are huge emerging issues related to shrinking critical dimension and next generation device structure, such as 3D DRAM, 3D V-NAND, 3D FET (gate-all-around nanowire and nanosheet FET) [23,24,25]. It is true that even they (current leading semiconductor device and equipment companies) do not have a clear alternative for these issues. These can be solved via modern LTP techniques [26,27,28], such as high aspect ratio & atomic layer plasma surface interactions, EUV plasma source, etc.

As noted above, the US program should invigorate LTP research at the universities and national labs into a sustained program to develop plasma manufacturing techniques and understanding of the plasma conditions and complex plasma-surface interactions that are the basis of these manufacturing techniques. This joint partnership should not exist just for the short period of time needed to meet the immediate international challenge but should be sustained to continue to promote innovations and maintained our leadership of the field. The program needs to be on a scale of $100's M/year funding for the national labs and universities. The universities are well suited to develop these new processing techniques and for the workforce development. The national labs are well suited to characterize and evaluate LTP manufacturing equipment for industry and provide a comprehensive study of the plasmas and complex plasma-surface interactions that are the basis of the manufacturing techniques. This initiative should not be funded at the expense of other plasma applications, fusion energy research (e.g. domestic program and ITER).

**US Leadership and Global Context**

The initiative is focused upon the US achieving the lead in nanomaterial plasma synthesis and plasma medicine. These are emerging fields that have many opportunities, but there is a lack of fundamental understanding of the processes.

Most importantly, the initiative is critical for the U.S. microelectronics industry. It focuses on the US regaining its lead in production of semiconductors from South Korea. In addition, the initiative will lead the way in the development of the next generation of computer chips based upon lithography at 10 nm and component sizes approaching 1 nm. The initiative is in direct response to the Made in China 2025 economic initiative. The US needs to support LTP research in the same manner it supports the medical and pharmaceutical fields with NIF funding, Exascale computing with its own initiative, and material science research with BES funding. Without this initiative and focus on Low Temperature Plasma the U.S. is at risk of losing its leadership position in microelectronics and not being competitive [29] in the developing fields of nanomaterial plasma synthesis and plasma medicine.

**Timeline of the Initiative**
(1) 1-2 years to form a strategy with consensus from the community and agreement with other related agencies
(2) 5-10 years to develop and establish this initiative

**Equipment/Facility Design Details**
Equipment design and manufacturing are completed by industry and most equipment can be purchased. There is no need for extensive facility designs.

**Cost Range**
$100 M per year to be impactful

**Cross-Cutting Connections**
This initiative has close relations with the following cross-cutting areas: Workforce development, Measurement & Diagnostic, Theory & Computation, Enabling technology.

**Advocates of This Initiative**
- Philip Efthimion, Yevgeny Raitses, and Igor Kaganovich
  *Princeton Plasma Physics Laboratory, Princeton University, Princeton, NJ 08543*
- Roberto Car and Mikhail Shneider
  *Princeton University, Princeton NJ 08540*
- Michael Keidar
  *George Washington University, Washington DC 20052*
- Hyo-Chang Lee
  *Korea Research Institute of Standards and Science, Daejeon 34113, Republic of Korea*

**Appendix**
International Technology Roadmap for Semiconductors 2.0: 2015 executive report (http://www.itrs2.net/) shows predictions for future NAND Flash parameters.

| NAND Flash | | | | | | | |
|---|---|---|---|---|---|---|---|
| Year of Production | 2015 | 2016 | 2020 | 2022 | 2024 | 2028 | 2030 |
| 2D NAND Flash uncontacted poly 1/2 pitch – F (nm) | 15 | 14 | 12 | 12 | 12 | 12 | 12 |
| 3D NAND minimum array 1/2 pitch -F (nm) | 80nm | 80nm | 80nm | 80nm | 80nm | 80nm | 80nm |
| Number of word lines in one 3D NAND string | 32 | 32-48 | 64-96 | 96-128 | 128-192 | 256-384 | 384-512 |
| Dominant Cell type (FG, CT, 3D, etc.) | FG/CT/3D | FG/CT/3D | FG/CT/3D | FG/CT/3D | FG/CT/3D | FG/CT/3D | FG/CT/3D |
| Product highest density (2D or 3D) | 256G | 384G | 768G | 1T | 1.5T | 3T | 4T |
| 3D NAND number of memory layers | 32 | 32-48 | 64-96 | 96-128 | 128-192 | 256-384 | 384-512 |
| Maximum number of bits per cell for 2D NAND | 3 | 3 | 3 | 3 | 3 | 3 | 3 |
| Maximum number of bits per cell for 3D NAND | 3 | 3 | 3 | 3 | 3 | 3 | 3 |

The stack number will reach 512 in 3D V NAND: extreme case of the high aspect ratio etch.

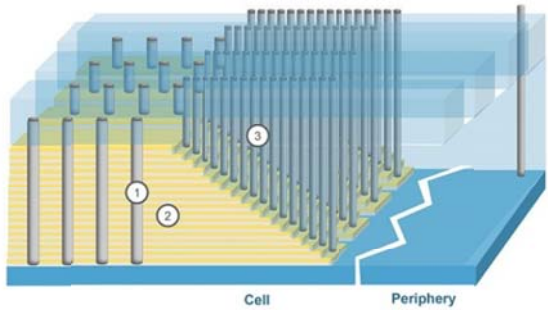
① Channel　②Gate Stack　③3D Shaping
3D V-NAND Flash memory structure of Samsung

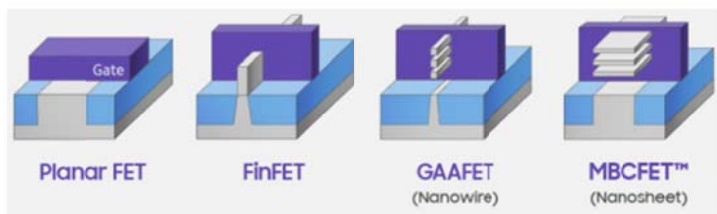

https://www.extremetech.com/computing/291507-samsung-unveils-3nm-gate-all-around-design-tools

Next logic device is gate-all-around (GAA) with nanowire and Multi-Bridge-Channel (MBC) with nanosheet FETs, which requires new challenging techniques (tailoring atomic layer plasma surface interactions techniques).